\documentclass[runningheads]{llncs}
\usepackage{graphicx}
\usepackage{subfig}
\usepackage[ruled,vlined, linesnumbered]{algorithm2e}
\usepackage{enumitem}

\newcommand{\quotes}[1]{``#1''}

\begin{document}
\title{Utilizing FastText for Venue Recommendation}
%
\titlerunning{Utilizing FastText for Venue Recommendation}
%
\author{Makbule Gulcin Ozsoy}
\authorrunning{MG Ozsoy}
%
\institute{University College Dublin\\
Dublin, Ireland\\
\email{makbule.ozsoy@ucd.ie}}
%
\maketitle              
\begin{abstract}
Venue recommendation systems model the past interactions (i.e., check-ins) of the users and recommend venues. Traditional recommendation systems employ collaborative filtering, content-based filtering or matrix factorization. Recently, vector space embedding and deep learning algorithms are also used for recommendation. In this work, I propose a method for recommending top-k venues by utilizing the sequentiality feature of check-ins and a recent vector space embedding method, namely the FastText. Our proposed method; forms groups of check-ins, learns the vector space representations of the venues and utilizes the learned embeddings to make venue recommendations. I measure the performance of the proposed method using a Foursquare check-in dataset.The results show that the proposed method performs better than the state-of-the-art methods.


\end{abstract}
\section{Introduction}\label{intro}
Location based social networks (LBSNs), e.g., FacebookPlaces, Foursquare, are used by millions of people who produce a vast amount of information through check-ins. The check-ins reveal not only the location preferences of the users but also the sequential, temporal and geographical information about the users and the venues. LBSN platforms use these kinds of information to improve their services. One of these services is venue recommendation systems.

Venue recommendation systems model the preferences of users on their past interactions (i.e., check-ins) and recommend venues that the users are more probable to visit. Traditional recommendation methods employ collaborative filtering, content based filtering and matrix factorization techniques. Recently, deep learning and word/document embedding methods gain more attention from the researchers. The embedding techniques, Word2Vec \cite{MikolovSCCD13} and Doc2Vec \cite{LeM14} are frequently used for making recommendations, e.g. \cite{ozsoy2016word}, \cite{manotumruksa2016modelling}, \cite{zhao2016gt}, \cite{yang2018unsupervised}. Although these techniques are powerful at learning the semantic relations among venues and users, a newer method named FastText \cite{bojanowski2016enriching} can be more performant to represent the venues. The FastText method extends the Word2Vec by representing each word as a bag of character n-grams. Therefore, it can represent almost any input, even if the same item is not encountered in the training.

Venue recommendation systems can use contextual features (e.g., sequentiality, temporality and spatiality) while predicting the future check-ins of the users. While the temporality and spatiality features are frequently used in venue recommendation literature, e.g. \cite{ZhangWang:2015}, \cite{HeLLSC16}, \cite{yang2018unsupervised}; sequentiality of check-ins gained limited attention, e.g. \cite{zhang2014lore}, \cite{zhao2016gt}. Guo et al.~\cite{guo2018exploiting} highlights that items used sequentially in a short time have a strong relationship with each other and capturing this information is important to make the right recommendations.

In this work, I utilize the FastText method to make venue recommendations while taking the sequentiality of check-ins into account. The contributions of this work are as follow:
\begin{itemize}
\item I use sequentiality of the check-ins and group the related check-ins together. Inspired by \cite{guo2018exploiting}, I consider the consecutive check-ins made in a short time by the same user as related. 
\item I utilize the FastText method from natural language processing (NLP) domain to learn the vector space representations of venues. The representations capture the regularities and semantic information of the venues. 
\item I use the learned vector space representations to make recommendations by considering the similarities among the vectors of venues. 
\item I use a Foursquare check-in dataset for the evaluation. I measure and compare the performance of the proposed method to the state-of-the-art methods. 
\end{itemize}

The paper is structured as follows: The review of the related work is given in the Section \ref{relWork}. The proposed method is explained in the Section \ref{appDeepRec}. The evaluation results are presented in the Section \ref{eval}. The paper is concluded in the Section \ref{conclusion}.

\section{Related Work}\label{relWork}
Recommendation systems estimate the preferences of users and suggest items based on the estimated preferences (\hspace{1sp}\cite{MassaA07}, \cite{TavakolifardA12}). Various approaches can be employed to make recommendations, from traditional methods like collaborative filtering, content based filtering, matrix factorization to more recent vector space embeddings, deep learning based methods. 

Collaborative filtering and content-based filtering methods use item-user or user-user similarities. Example collaborative filtering based venue recommendation methods belong to Ye et al.~\cite{Ye2010}, Yuan et al.~\cite{Yuan:2013:TPR:2484028.2484030}, Zhang and Wang\cite{ZhangWang:2015} and Ozsoy et al.~\cite{Ozsoy14}. Matrix factorization methods use the low-rank approximation of input data \cite{Ma:2011:RSS:1935826.1935877}. Example matrix factorization based recommendation methods belong to Pan et al.~\cite{pan2008one}, Hu et al.~\cite{hu2008collaborative}, Rendle et al.~\cite{rendle2009bpr}, Gao et al.~\cite{gao2013exploring}, Zhang et al.~\cite{zhang2014lore}, Li et al.~\cite{li2015rank}, Zhao et al\cite{ZhaoZYLK16} and He et al.~\cite{HeLLSC16}. None of the above-mentioned algorithms employ deep learning or vector space embeddings techniques. Recently these techniques have gained more attention from the recommendation systems domain. The example works utilizing deep learning for recommendation belong to Salakhutdinov et al.~\cite{SalakhutdinovMH07}, Georgiev and Nakov\cite{GeorgievN13}, Wang et al.~\cite{WangWY14}, Musto et al.~\cite{musto2018deep} and Ai et al.~\cite{ai2018learning}. These methods employ deep learning methods to make recommendations of movies, products etc., but not venues. 

Vector space embedding techniques from NLP are also used for making recommendations. Initial methods utilizing Word2Vec\cite{MikolovSCCD13} and Doc2Vec\cite{LeM14} for recommendation usually use text based features; e.g. tags, comments. Shin et al.~\cite{ShinCL14} employed Word2Vec to compute the vectors of tags in Tumblr and recommended Tumblr blogs to the users. Musto et al.~\cite{MustoSGL15} used the textual data collected from Wikipedia and employed Word2Vec to make movie recommendations. Recently, non-textual data is also used while learning the vector space embeddings for recommendation. Grbovic et al.~\cite{grbovic2015commerce} employed Word2Vec to predict the next purchase item. They treated purchase history of a user as the sentence and each product as the word. Guo et al.~\cite{guo2018exploiting} made product recommendations by employing a network embedding technique and collaborative filtering. They utilized the sequentiality of the items to identify related items. Even though these methods used vector space embedding techniques for recommendation, they did not focus on venue recommendation.

There are venue recommendation systems which use word/document embedding techniques in the literature. Ozsoy et al.~\cite{ozsoy2016word} made an analogy in between \quotes{sentences and all check-ins per user} and \quotes{words and individual check-ins} and employed Doc2Vec to make venue recommendations. Manotumruksa et al.~\cite{manotumruksa2016modelling} made venue recommendations by inferring the vector space representations of venues. They utilized the textual content of the comments to model the preferences of users and the characteristics of venues. Liu et al.~\cite{liu2016exploring} employed Skip-Gram (a Word2Vec technique) and C-WARP loss in their method SG-CWARP to learn the latent representation of the users and the items. The learned representations are used for making location recommendations. Zhao et al.~\cite{zhao2016gt} used Word2Vec to learn the venue embeddings in their method SEER. Additionally, they incorporated temporal (T-SEER) and geographical information (GT-SEER). They expanded their proposed method in \cite{zhao2017geo}. Yang and Eickhoff\cite{yang2018unsupervised} made location recommendations by utilizing Word2Vec techniques in their STES algorithm. They incorporated geographical, temporal and categorical information to model places, neighborhoods and users.

Many researchers exploit the usage of embeddings and contextual information for venue recommendation. To our knowledge, none of them has employed the FastText method \cite{bojanowski2016enriching} and only a few of them utilized the sequentiality of check-ins, e.g. \cite{zhang2014lore}, Zhao et al.~\cite{zhao2016gt}. In this work, I use the sequentiality of the check-ins and employ the FastText method to make venue recommendations.

\section{Utilizing FastText to Recommend Venues}\label{appDeepRec}
I propose a method to recommend top-k venues. The method groups the check-ins by utilizing the sequentiality of check-ins, learns the vector space embeddings of venues by the FastText method \cite{bojanowski2016enriching} and uses the learned embeddings to make recommendations. 

\subsection{The FastText method}\label{fastText}
Vector space embedding models represent the documents, words and subword units as vectors to capture the contextual and semantic relations among these textual units \cite{MikolovCoRR13}. A recently developed vector space embedding method, namely the FastText method \cite{bojanowski2016enriching}, extends the Word2Vec \cite{MikolovSCCD13} by representing each word as a bag of character n-grams. 

Given the input sentences (sequence of words), Word2Vec captures the semantic and syntactic information of the words and produces low dimensional continuous space representations of them \cite{LiXTJZC15}. The two techniques of the Word2Vec, namely Skip-Gram and CBOW, use the words as they appear on the input, i.e., without any morphological analysis, and produce different vectors for words even if they share common roots. This can become problematic for rare words, morphologically rich languages; e.g. Turkish;  inflected languages; e.g. Spanish; or languages with compound words; e.g. German. Word2Vec learns only the representation of the words existing in the training data. In the execution time, if an unseen word; even an inflected or compound word; is encountered, Word2Vec cannot return any representation for that word. For example, when the Spanish word \quotes{corres} (to run) is in the training set but the word \quotes{corro}(I run) is not, the Word2Vec techniques cannot return any output for the word \quotes{corro}. In order to overcome the limitations of Word2Vec, Bojanowski et al.~\cite{bojanowski2016enriching} proposed the FastText method which extends the Word2Vec and takes the subword units (character n-grams) into account.

\begin{figure}[!bt]
\begin{minipage}{1.0\linewidth}
\centering
     \subfloat[Vector representations for input subwords\label{exampleFastTextTrain}]{%
       \includegraphics[width=0.7\textwidth]{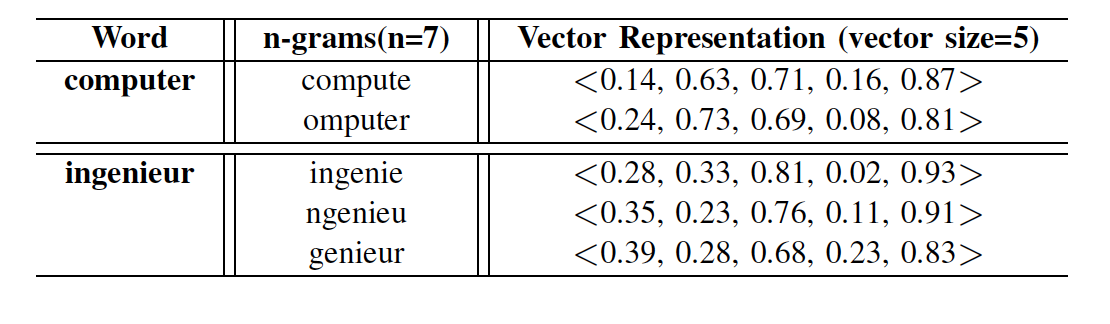}
     }
     \hfill
     \subfloat[Vector representations for an unseen word\label{exampleFastTextTest}]{%
       \includegraphics[width=0.7\textwidth]{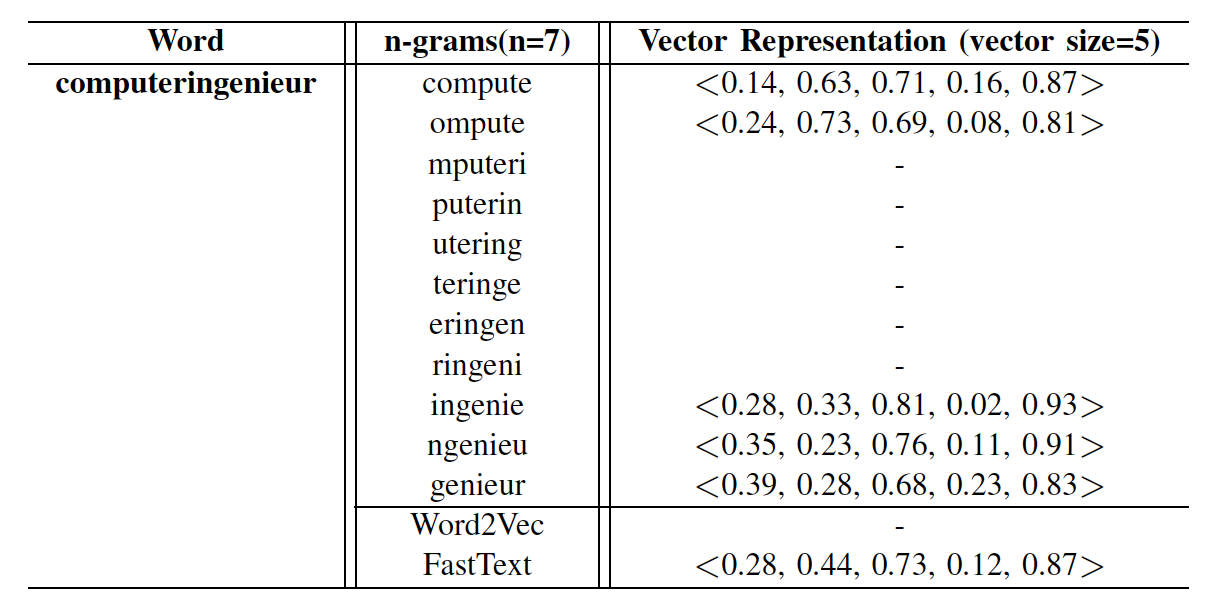}
     }
\end{minipage}

\caption{Examples describing how the FastText method learns the vector representations of words and subwords}
\label{exampleFastTextForText}
\end{figure}

In the training time, the FastText method models each input word as a bag of character n-grams and produces the vector representations of these n-grams as well as the input words. The output vector representation of the input words are calculated by combining the learned vectors of the word itself and its n-grams. In the execution time, the queried word can be either already seen in the training data or unseen/new word. In the former case, the output vector representation of the word is directly returned. In the latter case, the FastText method calculates the vector of the queried word by taking the average of vectors of its character n-grams. For example, assume that, the queried word is \quotes{computeringenieur} (computer engineer) and it is not seen while training the model. However, the words \quotes{computer} (computer) and \quotes{ingenieur} (engineer) are in the training data and the vector representations of their character n-grams ($n=7$) are learned and shown in Figure \ref{exampleFastTextTrain}. The FastText method approximates the unseen word's (computeringenieur) vector representation from the vectors of its n-grams, as shown in Figure \ref{exampleFastTextTest}. 
 



\subsection{Recommendation using the FastText method} \label{fastTextCheckin}
The proposed method has three steps: 1) Representing the check-ins as groups by using the sequentiality of the check-ins 2) Learning the vector space embeddings of the venues using the FastText method 3) Using the learned vector representations to recommend venues. 

\begin{table}[!bt]
\SetAlFnt{\small}
\SetAlCapFnt{\small}
\SetAlCapNameFnt{\small}
\caption{Our proposed analogy}\label{analogy}
\centering
\begin{tabular}{l||l}
\hline
Natural Language Processing (NLP) & Recommendation Systems (RS)\\
\hline
sentences			&	 all check-ins per user \\
words 		&	 groups (sequences) of check-ins\\
sub-words (character n-grams) 		&	sub-sequences of check-ins \\
characters  	& individual check-ins\\
\hline
\end{tabular}
\end{table}

\begin{figure}[!bt]
\begin{minipage}{1.0\linewidth}
\centering
     \subfloat[Sentence, words and subword units\label{fastTextExampleSentence1}]{%
       \includegraphics[width=0.35\textwidth]{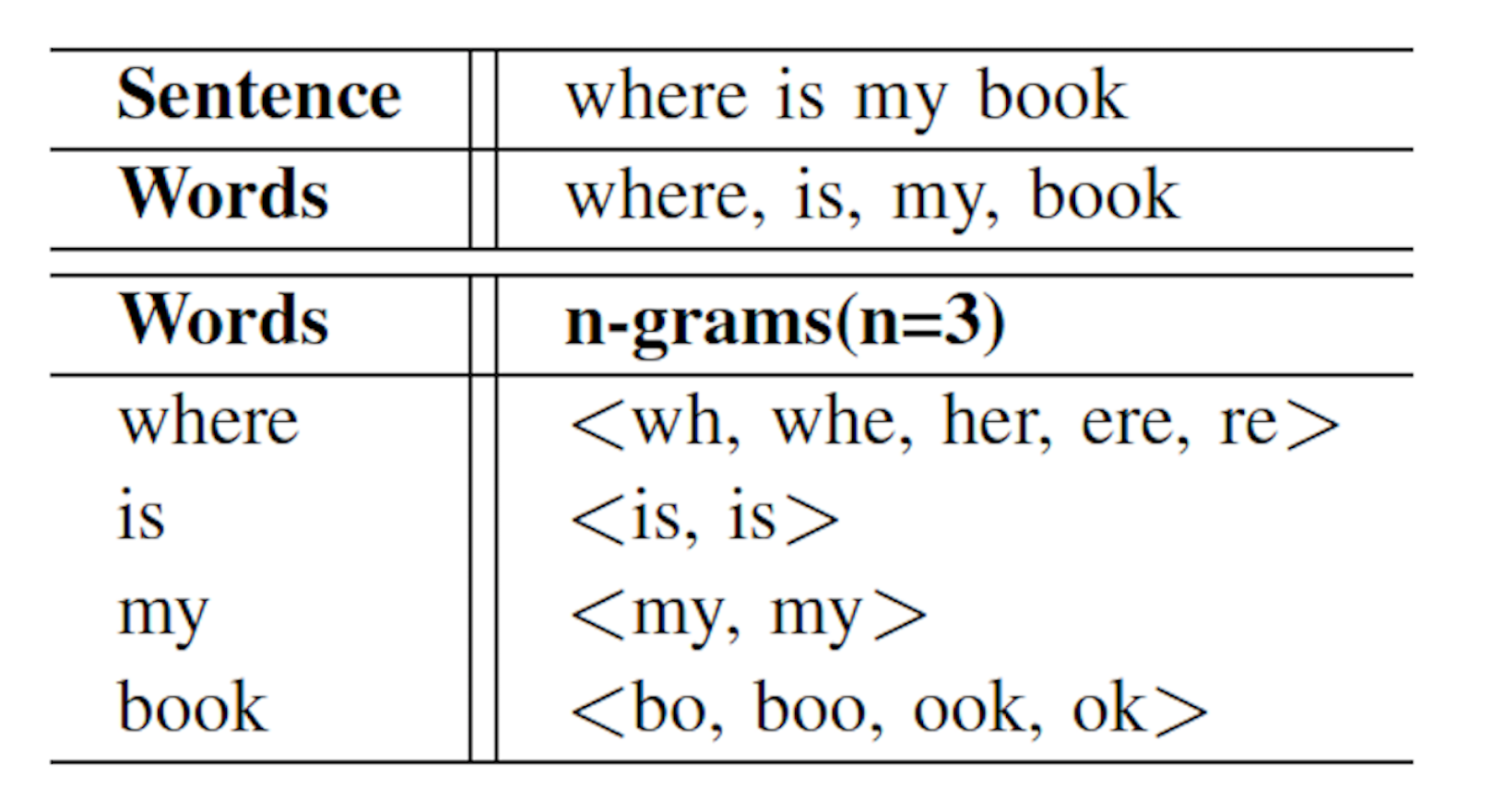}
     }
     \hfill
     \subfloat[All check-ins of a single user, sequence of check-ins and sub-sequence of check-ins\label{fastTextExampleSentence2}]{%
       \includegraphics[width=0.59\textwidth]{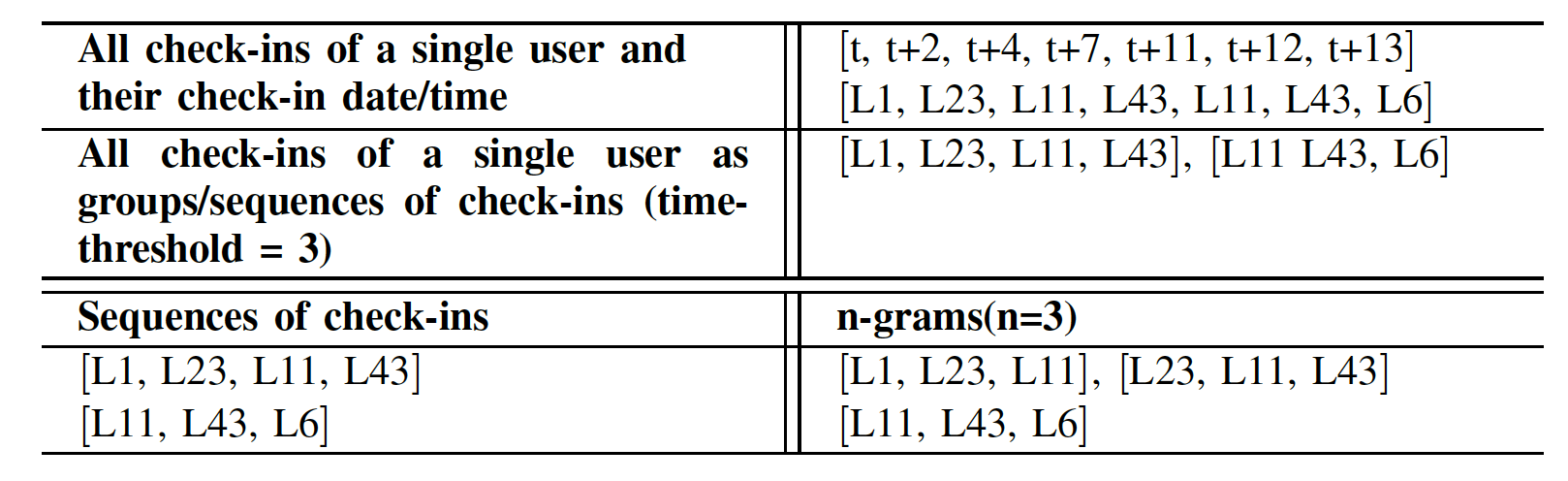}
     }
\end{minipage}\par\medskip

 \caption{An example describing our analogy}
\label{fastTextExampleSentences}
\end{figure}

While utilizing FastText for recommendation, I make an analogy in between textual data and check-ins. The proposed analogy is in between: \quotes{sentence and all check-ins per user}, \quotes{words and groups/sequences of check-ins}, \quotes{subword units (character n-grams) and sub-sequences of check-ins} and \quotes{characters and individual check-ins} (Table \ref{analogy}). Figure \ref{fastTextExampleSentences} exemplifies our analogy. In Figure \ref{fastTextExampleSentence1}, the input sentence (\quotes{where is my book}) is split into its words([\quotes{where}, \quotes{is}, \quotes{my}, \quotes{book}]) and then the subword units are formed by using character n-grams ($n=3$). Figure \ref{fastTextExampleSentence2} presents the corresponding elements of venue recommendation systems to the NLP domain (sentences, words and subword-units). In the figure, the example user checks in at multiple locations $L$ in different dates/times $t$. Those locations are split into groups/sequences of check-ins, which then form sub-sequences of check-ins by using n-grams of individual check-ins. 

\subsubsection{Representing the check-ins as sequences and sub-sequences} \label{fastTextCheckin_grouping}
The first step of our proposed method is to represent the check-ins as groups/sequences and sub-sequences by utilizing the sequentiality feature of the check-ins. The process is presented in the Algorithm \ref{fastTextCheckin_algo}. 

\begin{algorithm}[!bt]
\SetAlFnt{\small}
\SetAlCapFnt{\small}
\SetAlCapNameFnt{\small}
\KwIn{List of check-ins of a single user ($L$)}
\KwOut{Sub-sequences of check-ins ($subseq_L$)}
	 Split $L$ into groups/sequences of check-ins $seq_L$\\
	 Form sub-sequences of check-ins $subseq_L$ from $seq_L$ (using n-grams)\\
\Return $subseq_L$
\caption{{\bf Producing sub-sequences of input check-ins} \label{fastTextCheckin_algo}}
\end{algorithm}

The first step of the algorithm forms the groups/sequences from all check-ins of a user. This is analogical to splitting the sentences into words. Extracting words from sentences is intuitive, e.g. in English I split the input sentence by the spaces to find the words. However, it is less intuitive to split the check-ins and form the groups/sequences. For this purpose, I employ a technique inspired by \cite{guo2018exploiting}. Guo et al.~\cite{guo2018exploiting} states that the items which are rated in a short time interval are more likely to be correlated. In order to decide the correlation between two items, they use time difference between ratings. Similar to their technique, I use the time of check-ins and their sequentiality to form the groups of check-ins: \textit{Given the ordered list of all check-ins ($C_u$) of a single user $u$  with the time of the check-ins ($t_i$); i.e., $ C_u = \{(c_0, t_0), (c_1, t_1),  ...., (c_n, t_n) ; t_0 \leq t_1 \leq ... \leq t_n\}$; all the related check-ins are grouped together. Two consecutive check-ins $(c_i, t_i)$ and $(c_j, t_j)$ are considered as related if they are made by the same user and $0 \leq t_j - t_i \leq \Delta T$ where $\Delta T$ represents a short time interval.}


In Figure \ref{fastTextExampleSentence2}, for example, having the list of timestamps of check-ins and $\Delta T=3$, two groups of check-ins are formed. The split occurs in between two consecutive the check-ins whose timestamps are $t+7$ and $t+11$, because the time between these check-ins is more than the assigned $\Delta T$; i.e., $(11-7) > 3$.

The second step of the Algorithm \ref{fastTextCheckin_algo} extracts the sub-sequence of check-ins from the groups/sequences, which is analogical to extracting n-grams of the words. For both words and groups/sequences of check-ins, I apply the same process: When I have a word as the input, I form the n-grams from its characters. When I have a group of check-ins as the input, I form the n-grams from the individual check-ins in that group. 

\subsubsection{Learning the vector space embeddings of the venues} \label{fastTextCheckin_embedding}
\begin{figure}[!bt]

\begin{minipage}{1.0\linewidth}
\centering
     \subfloat[Vector representations for input sub-sequence of check-ins\label{fastTextExampleLearrntCheckins}]{%
       \includegraphics[width=0.7\textwidth]{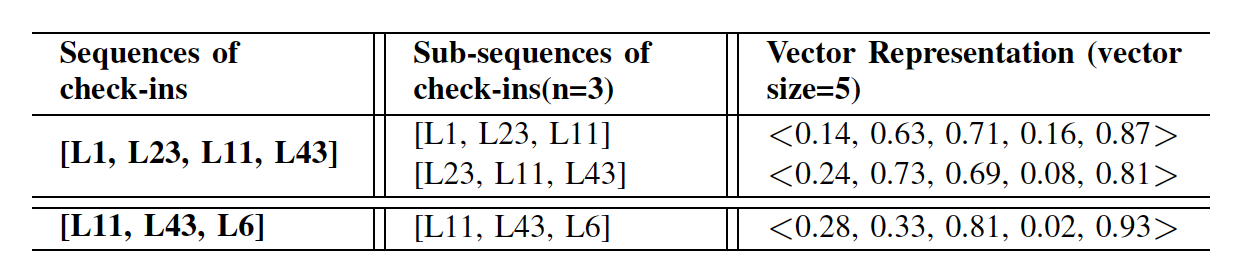}
     }
     \hfill
     \subfloat[Vector representations for an unseen sequence of check-ins\label{fastTextExampleUnseenUserCheckins}]{%
       \includegraphics[width=0.7\textwidth]{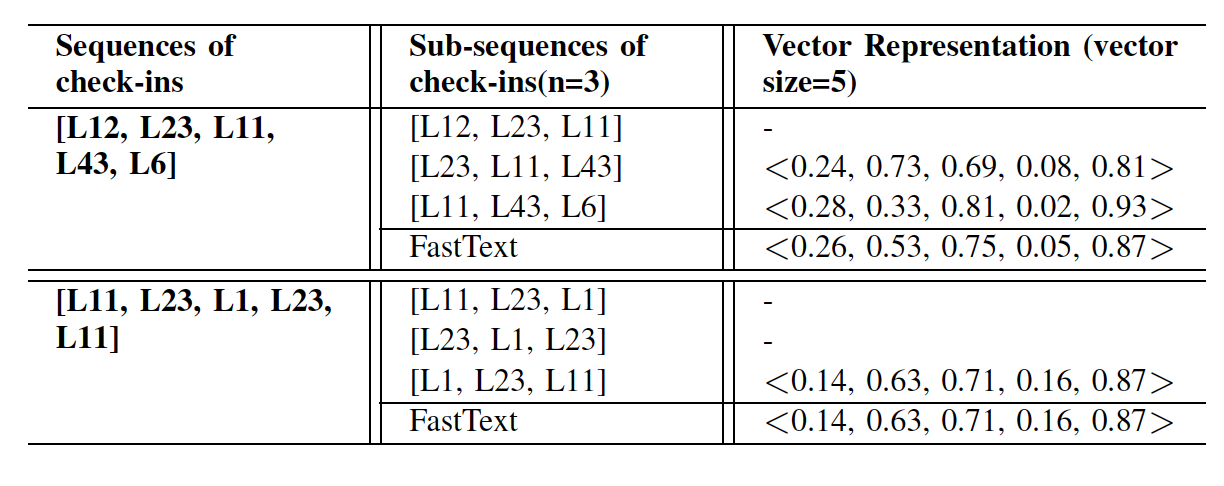}
     }
\end{minipage}
\caption{Examples describing how the FastText method learns the vector representations of venues}
\label{exampleFastTextForCheckin}
\end{figure}

The second step of our proposed method is to learn the vector space embeddings of the venues by utilizing the FastText method. The process is presented in the Algorithm \ref{fastTextCheckin_embedding_algo}. Firstly, the sub-sequences of check-ins are extracted. Then, the model is trained using the FastText method by considering all the extracted sub-sequences of check-ins from all of the users. In our implementation, I used the FastText implementation of the \textit{gensim} toolbox \cite{rehurek_lrec} and modified it according to our needs\footnote{The FastText implementation in gensim extracts the n-grams during the learning process. Since our definition of n-grams is slightly different than extracting n-grams from words, I modified the related code. The code is on Github: https://github.com/mgulcin/FastTextRec/}. The result of training with FastText is the vector representations of the input sequences and the sub-sequences. In Figure \ref{fastTextExampleLearrntCheckins}, I present example learned vectors of sequence and sub-sequence of check-ins. 

\begin{algorithm}[!bt]
\SetAlFnt{\small}
\SetAlCapFnt{\small}
\SetAlCapNameFnt{\small}
\KwIn{List of all check-ins per user ($L_{All}$)}
\KwOut{Vector space embeddings of sub-sequences of check-ins}
$subseq\_L_{All}=\{\}$ \\ 
\ForEach{all check-ins of a single user $L$ in  $L_{All}$}{%
 	 Produce the sub-sequences of check-ins $subseq_L$  from the input check-ins $L$ (The Algorithm \ref{fastTextCheckin_algo})\\
	 Collect the extracted $subseq_L$ in $subseq\_L_{All}$} 
Learn the vector space embeddings using the values of $subseq\_L_{All}$\\
\Return Vector space embeddings of the sub-sequences of check-ins ($V$)
\caption{{\bf Learning the vector space embeddings of venues} \label{fastTextCheckin_embedding_algo}}
\end{algorithm}

\subsubsection{Making venue recommendations using the vector space embeddings}\label{fastTextCheckin_recommendation}
The last step of our proposed method is to make recommendations by 1) Calculating the similarity among the venues and 2) Recommending the most similar venues to the previous check-ins. This approach is similar to the traditional content based filtering method, which uses the predefined features describing the items and users. In our proposed method, instead of predefined features the vector representations learned by the FastText method are used. 

The process is presented in the Algorithm \ref{fastTextCheckin_recommendation_algo}. The algorithm, firstly, extracts the vector space embeddings of the input check-ins. For this purpose, the groups/sequences of the check-ins are formed. The groups can be either already seen or unseen in the training data. In the former case, the vector of the group/sequence, which is learnt by FastText, is directly used. In the latter case, the vector of the group/sequence is calculated using the known vectors of the sub-sequences. For example, assuming the vectors of the groups/sequences in Figure \ref{fastTextExampleLearrntCheckins} are already learned, when requested their vector representations is directly returned. However, when the groups/sequences presented in Figure \ref{fastTextExampleUnseenUserCheckins} which are unseen in the training are requested, their vectors are calculated using the known vectors of their sub-sequences. 
 
\begin{algorithm}[!bt]
\SetAlFnt{\small}
\SetAlCapFnt{\small}
\SetAlCapNameFnt{\small}
\KwIn{Vector space embeddings of the sub-sequences of check-ins ($V$), List of previous check-ins per user ($L_{All}$), Recommendation type  ($rec_{type}$)}
\KwOut{List of recommendations per user}
$user\_2\_recommendations=\{\}$ \\ 
\ForEach{previous check-ins of a single user $L$ in  $L_{All}$}{%
 	Extract the vector space embeddings of the input check-ins $L$\\
	Find most similar venues to $subseq_L$ and their similarity scores using the vector space embeddings $V$\\
	Decide on the top-k recommendations $rec$ based on recommendation type $rec_{type}$\\
	Collect the recommendations $rec$ in $user\_2\_recommendations$} 
\Return $user\_2\_recommendations$
\caption{{\bf Making venue recommendations using the vector space embeddings} \label{fastTextCheckin_recommendation_algo}}
\end{algorithm}

Secondly, the Algorithm \ref{fastTextCheckin_recommendation_algo} calculates the similarity among vectors and finds the most similar N venues (neighbors) to the previously visited venues. For the similarity calculations, I used Cosine Similarity\footnote{I utilized $most\_similar$ method provided by the \textit{gensim} toolbox, which calculates Cosine Similarity among the vectors and returns the most similar N vectors to the target vector.}. The output of this step provides the most similar sequence of check-ins (i.e., a sequence of venues to visit), not the individual venues, because in our proposed method the learned vectors actually represent a sequence not individual venues. However, the traditional recommendation methods, e.g. content based recommendation, make recommendations of individual venues. In order to make recommendations of individual venues, I added a post-processing step. In this last step, the Algorithm \ref{fastTextCheckin_recommendation_algo} decides the top-k venue recommendations using the calculated similarity scores and the input recommendation type, which are proposed in this work. I defined three recommendation types: 

\begin{itemize}
\item Sequence of Venues (Seq): This approach directly recommends a group of venues based on the similarity scores. For example, for the check-ins shown in Figure \ref{fastTextExampleUnseenUserCheckins} and learned vectors shown Figure \ref{fastTextExampleLearrntCheckins}, lets assume that the most similar 2 ($N=2$) groups and their similarity scores are: (\textit{[L1, L23, L11, L43], 0.99}), (\textit{[L11, L43, L6], 0.97}). Then, the Seq recommendation type will recommend $[L1, L23, L11, L43]$, because it has the highest similarity score. 
\item Individual Venues: This approach calculates a similarity score per venue: 1) Assign the same similarity score to each venue in a group and create a list of similarity scores per venue 2) Calculate the overall similarity score per venue. 3) Recommend the top-k most similar venues. The second step can be performed in several different ways. I utilized two versions:
\begin{itemize}
\item Seq-Single-Max: The maximum of the similarity scores is assigned as the overall score. For the similarity scores provided above, the score of each venue is $L1=0.99, L23=0.99, L11=0.99, L43=0.99, L6=0.97$ and the recommendation can be $[L1, L23]$ if $k=2$. 
\item Seq-Single-Avg: The average of the similarity scores is assigned as the overall score. For the example presented above, the list of similarities of $L23$ is $[0.99, 0.97]$ and its overall similarity score is $0.98$.  
\end{itemize}
\end{itemize}
 
\section{Evaluation}\label{eval}
I present the dataset, the evaluation metrics, the parameters and the evaluation results in the following sections.

\subsection{Dataset, Evaluation Metrics and Parameters}\label{dataset_evalMetrics_expParameters}
I use a subset of the Checkin2011 dataset \cite{GaoTL12} for the evaluation. The original dataset is collected from Foursquare between Jan. 2011 - Dec. 2011 and contains 11326 users, 187218 venues and 1385223 check-ins.  Zhao et al.~\cite{zhao2016gt} used a subset of this dataset with some pre-processing. I followed their approach: 1) The subset of the dataset is extracted by limiting the time period to Jan. 2011 - Aug. 2011. 2) The venues visited less than five times are filtered out. 3) The users with less than or equal to ten check-ins are filtered out. The resulting dataset contains 10034 users, 16561 venues and 865647 check-ins. For the evaluation, the first $80\%$ of each user's check-ins is used for training and the rest is used for testing; i.e., the ratios of training and test sets are $80\%$ and $20\%$.

The performance is evaluated by the Precision@k, NDCG and HitRate metrics. The Precision@k measures the relevance of items on the output list. The NDCG@k decides the relevance of the listed items depending on their rank. HitRate measures the ratio of users who are given at least one true recommendation and it is calculated by: $HitRate = \frac{\sum_{m \in M} HitRate_m}{|M|}$. In the equation, $m$ is the individual user and $M$ is the total set of users. $HitRate_m$ is equal to $1.0$ if there is at least one true recommendation for the user $m$ and $0.0$ otherwise. While giving the evaluation results, the performance metrics are calculated separately for each user and then the overall averages per metric are presented.

Each step of our proposed method contains different parameters. In order to form the groups of check-ins, $\Delta T$ value is used. Its value for our dataset is calculated by 1) Measuring the time between the consecutive check-ins per user per day. 2) Calculating the mean and standard deviation of these values for each user. On the average, the time between two check-ins is found to be around 18500 seconds ($\sim$5 hours). Using this observation, I decided to consider two check-ins which are made in less than 5 hours are related, i.e., I assigned  $\Delta T = 5$ hours. 

In order to learn the vector representations of venues, I used the FastText implementation in $gensim$ toolbox, which provides several parameters. I experimented on three parameters and used the default values for the others\footnote{Except setting $min\_count=1$ and $min\_n=1$. $min\_count$: Minimum word frequency. If the word's frequency is less than $min\_count$, it is ignored. $min\_n$: Minimum length of char n-grams to be used for training.}:
\begin{itemize}
\item $sg$: Type of the training algorithm (Skip-gram or CBOW). I used both algorithms for the experiments. 
\item $max\_n$: Maximum length of character n-grams. I used values in the range $[1,10]$ with increment of 1 (Keeping $size=100$).
\item $size (s)$: Size of the word vectors. I experimented using the following values: $[10, 50, 100, 150, 200, 250]$ (Keeping $max\_n=5$).
\end{itemize}

In order to make recommendations by using learned vectors, I used two parameters: Number of neighbors ($N$) and output list size ($k$). I set both $N$ and $k$ to 10 for the experiments. 

\subsection{Evaluation Results}\label{evalResults}
I evaluate the proposed method in three folds: 1) The effect of the parameters and recommendation types 2) The effect of utilizing the FastText method and explicitly grouping check-ins 3) Comparison to the state-of-the-art methods.

\subsubsection{Effect of different parameters and recommendation types}\label{evalParams}
I experimented on three parameters, namely $sg$,  $max\_n$ and $size (s)$, which are explained in the Section \ref{dataset_evalMetrics_expParameters}. For each recommendation type, the best performing results per evaluation metric are selected and presented together with their parameter settings in Table \ref{fastTextResults}.

\begin{table}[!bt]
\SetAlFnt{\small}
\SetAlCapFnt{\small}
\SetAlCapNameFnt{\small}
\caption{Evaluation results for FastText based methods}\label{fastTextResults}
\centering
\begin{tabular}{l||c||c||c||c}
\hline
\textbf{Method} & \textbf{Parameters} & \textbf{Precision} &  \textbf{NDCG} & \textbf{HitRate} \\
\hline
FastText-Seq & Skip-Gram, s=100, max\_n=3	&	0.0044 &  	0.0066	&  0.0419 \\
\hline
FastText-Seq-Single-Max & Skip-Gram, s=100, max\_n=9 &	0.0819 &  	0.2170	&  0.7164 \\
\cline{2-5}
 & Skip-Gram, s=100, max\_n=5 &	0.0812 &  \textbf{ 0.2179}	 	&  \textbf{0.7175} \\
\hline
FastText-Seq-Single-Avg & Skip-Gram, s=100, max\_n=9	 	&	\textbf{0.1040} &  	0.1554	&  0.6142 \\
\cline{2-5}
 & CBow, s=100, max\_n=8		&	0.1008 &  	0.1474	&  0.6192 \\
\hline
\end{tabular}
\end{table}

Table \ref{fastTextResults} shows that Skip-Gram mostly performs better than CBOW. This can be explained by the observation made by Mikolov et al.~\cite{MikolovSCCD13}, which states that the Skip-Gram is better at representing rare words/phrases and works well with small datasets, whereas CBOW is better at representing the frequent words and works better with large datasets. The Seq type performs worse than others. This is expected because making recommendation of sequences is more challenging. While deciding true positives, I strictly looked for exact matches. For example, if the actual and the recommended check-ins are $[L1, L2, L3]$ and $[L1, L2, L4]$, respectively; the recommendation is considered as wrong (false positive) even if the first two venues match. Table \ref{fastTextResults}, which presents only the the best performing settings, reveals that the vector size 100 works better than the other experimented values. 

\begin{figure}[!bt]
\begin{minipage}{1.0\linewidth}
\centering
     \subfloat[Different vector sizes ($size$) (where $max\_n=5$)\label{skipgram_vs}]{%
       \includegraphics[width=0.40\textwidth]{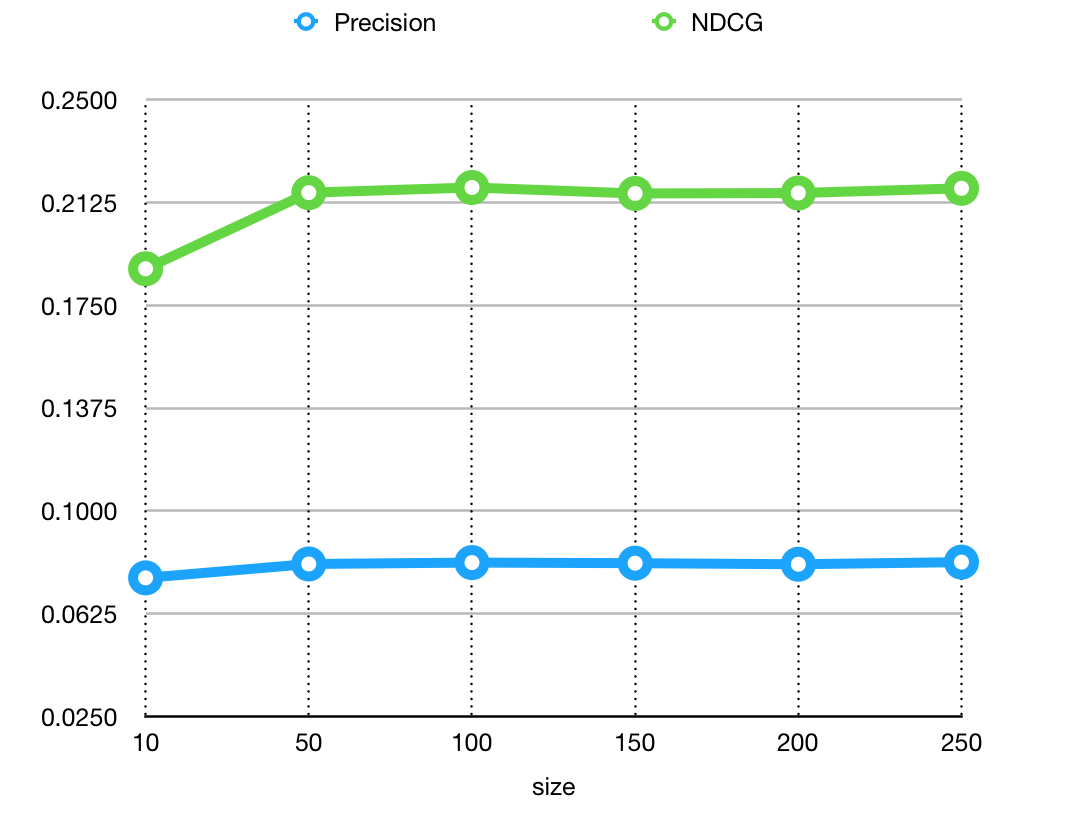}
     }
    \hfill
     \subfloat[Different length of character n-grams ($max\_n$) (where $size=100$)\label{skipgram_maxn}]{%
       \includegraphics[width=0.40\textwidth]{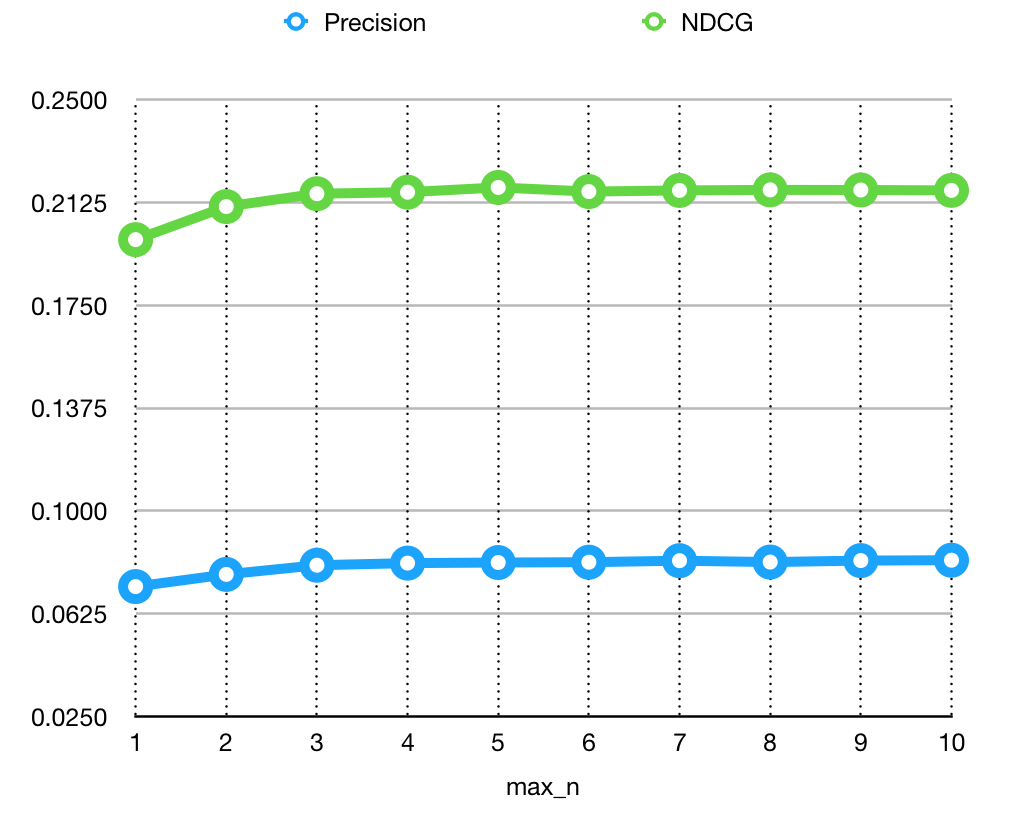}
     }
\end{minipage}\par\medskip

 \caption{Performance results of Skip-gram (Seq-Single-Max) for different vector sizes ($size$) and for different length of character n-grams ($max\_n$)}
\label{skipgram_perf}
\end{figure}

Figure \ref{skipgram_vs} presents the performance of the Skip-Gram (Seq-Single-Max) on different vector sizes\footnote{The observations for CBOW and other recommendation types (i.e. Seq, Seq-Single-Avg, Seq-First) are similar.}. The figure shows that the smaller vector size (e.g. $size=10$) does not work well, but the performance of larger vector sizes are close to each other. Figure \ref{skipgram_maxn} presents the performance of the Skip-Gram (Seq-Single-Max) on different lengths of n-grams. The figure reveals that the performance of our method does not change much with the length of n-grams ($max\_n$), but setting the $max\_n$ to a larger value than 1 works better. 

\subsubsection{Effect of  utilizing the FastText method and explicitly grouping check-ins}\label{evalSequential}
In order to measure the effectiveness of utilizing the FastText method rather than another vector space embedding method, I used a set of Word2Vec based methods. I changed the vector space representation learning step of our algorithm from FastText to Word2Vec. As a result, the algorithm uses the sequence of check-ins instead of sub-sequence of check-ins while learning the embeddings. Then the learned embeddings are used in exactly the same way as described before, in the Section \ref{fastTextCheckin_recommendation}. For the implementation I continued to use the $gensim$ toolbox by setting the $word\_ngrams$ parameter to 0\footnote{In $gensim$, setting the $word\_ngrams$ parameter to 0 makes the implementation equivalent to Word2Vec.}.

In order to measure the effectiveness of utilizing the sequentiality and grouping the check-ins explicitly, I defined an additional recommendation type; namely Non-Sequential (Non-Seq). It considers each individual check-in as the input, instead of using the group/sequence of check-ins. For example, for the input presented in Figure \ref{fastTextExampleUnseenUserCheckins}, each individual check-in (i.e., $L12, L23, L11, ...$) is considered as an input to the Word2Vec.


\begin{table}[!bt]
\SetAlFnt{\small}
\SetAlCapFnt{\small}
\SetAlCapNameFnt{\small}
\caption{Evaluation results for Word2Vec based methods}\label{word2VecResults}
\centering
\begin{tabular}{l||c||c||c||c}
\hline
\textbf{Method} & \textbf{Parameters} & \textbf{Precision} &  \textbf{NDCG} & \textbf{HitRate} \\
\hline
Word2Vec-NonSeq & CBow, s=150	&	0.0153	&	0.0219	&	0.1320\\
\hline
Word2Vec-Seq & 	Skip-Gram, s=100	& 0.0043		&	0.0065	&	0.0403 \\
\hline
Word2Vec-Seq-Single-Max& 	Skip-Gram, s=150 	&	0.0815	&	0.2161	&	\textbf{0.7115} \\
\cline{2-5}
& 	Skip-Gram, s=200  &	0.0812	&	\textbf{0.2164}	&	0.7109 \\
\cline{2-5}
& 	Skip-Gram, s=250	&	0.0811	&	0.2162	&	0.7153	\\
\hline
Word2Vec-Seq-Single-Avg & 	Skip-Gram, s=100		&	\textbf{0.1034}	&	0.1541	&	0.6106 \\
\cline{2-5}
& CBow, s=50					&	0.1012	&	0.1487	&	0.6125\\
\hline
\end{tabular}
\end{table}

The evaluation results of Word2Vec based methods are sightly worse than the results for the FastText based methods. However, considering different methods and parameters their behavior are similar (Table \ref{word2VecResults}): Skip-Gram usually performs better, the small vector sizes do not perform well and the Seq type performs worse compared to the other types. The best Precision, NDCG and HitRate scores are obtained by Word2Vec-Seq-Single-Max and Word2Vec-Seq-Single-Avg types and these scores are slightly worse than the performance of FastText based methods. The additional recommendation type (Non-Seq) performs worst among all the recommendation types. In overall, the comparison reveals that using the sequentiality of check-ins and grouping the check-ins explicitly, as in FastText-based methods, capture the relations among the venues better. 

\subsubsection{Comparison to the state-of-the-art methods}\label{evalCompare}
\begin{figure}[!bt]
\centering
\includegraphics[width=0.47\textwidth]{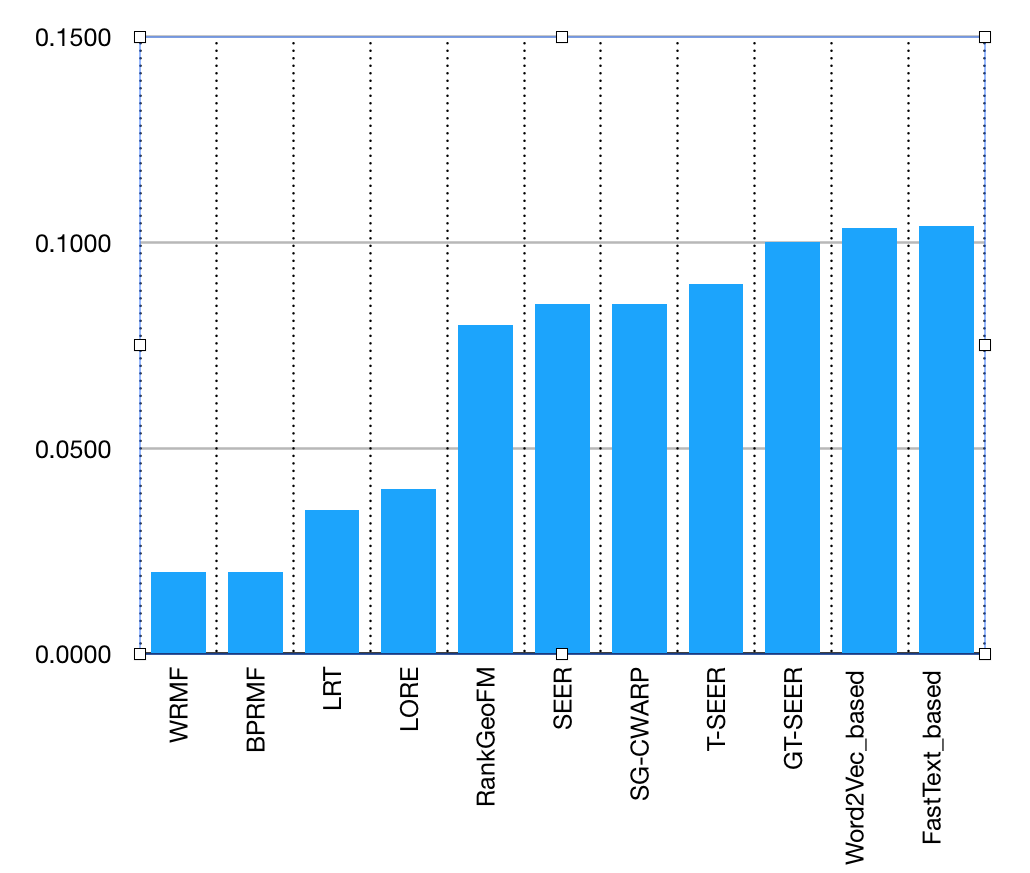}
\caption{Comparison with the state-of-the-art methods (Precision@10)}\label{compareResults}
\end{figure}

I compared our proposed method with the state-of-the-art methods (Figure \ref{compareResults}). For the comparison, I presented the results of the FastText-Seq-Single-Avg and Word2Vec-Seq-Single-Avg. In order to obtain the performance of the state-of-the-art methods, I referred to the evaluation results presented in \cite{zhao2016gt} and \cite{zhao2017geo}. They use exactly same dataset as ours and present evaluation results of the following methods: 1) Matrix factorization based methods: BPRMF, WRMF; 2) Non-embedding based methods which incorporate contextual information:  LRT,  LORE, Rank-GeoFM and 3) Embedding based methods: SG-CWARP, SEER, T-SEER and GT-SEER, where the latter two incorporate temporal and geographical information.

The comparison results reveal that our proposed method, FastText-Seq-Single-Avg, performs the best. The performance of Word2Vec-Seq-Single-Avg and GT-SEER are close to the FastText-Seq-Single-Avg. Both of these methods use Word2Vec in order to learn the venue embeddings. This reveals that the vector space embeddings are able to capture the relations among the venues. The other X-SEER methods (SEER and T-SEER) do not perform as well as GT-SEER. This shows that geographical features provides useful information to make venue recommendations. SG-CWARP which uses Word2Vec embedding technique performs worse than other embedding based techniques. The non-embedding based methods that I used for comparisons do not perform as well as embedding based methods. One exception is Rank-GeoFM. It incorporates geographical and temporal influence and its performance is close to the SEER but still worse than FastText-Seq-Single-Avg. These observations show that forming explicit groups/sequences of check-ins and utilizing FastText to learn vector representations of venues is useful to make venue recommendations.

\section{Conclusion}\label{conclusion}
I propose a method to recommend top-k venues. The method takes the sequentiality of check-ins into account and utilizes the FastText method from the NLP domain to learn the vector space embeddings of venues. The learned vector representations are used for calculating the similarity among venues and for recommending the most similar venues to the already visited ones. I execute the experiments on a Foursquare check-in dataset. The proposed method outperforms the state-of-the-art methods. The results reveal that forming explicit groups of check-ins and utilizing FastText is promising for making venue recommendations. 

In the future, I will incorporate geographical information to our FastText based recommendation method. I will also compare our method's performance on cold-start users with its performance on active users. Even though the focus of this work was making venue recommendations, I believe that the proposed method can be used for other recommendation problems, e.g. music recommendation. In the future I also want to make experiments on other datasets related to different recommendation problems.

%
%
%
\bibliographystyle{splncs04}
\bibliography{UtilizingFastText}
%




\end{document}